# Amplifying thermal conduction calibre of nanocolloids employing induced electrophoresis

**Purbarun Dhar** [a, *], **Lakshmi Sirisha Maganti** [b, #,] **A R Harikrishnan** [c, #] and **Chandan Rajput** [c, #]

[a] Department of Mechanical Engineering, Indian Institute of Technology Ropar, Rupnagar–140001, India

[b] Department of Mechanical Engineering, State University of New York Binghamton, Binghamton, NY, USA 13902

[c] Department of Mechanical Engineering, Indian Institute of Technology Madras, Chennai–600036, India

[#] Equal contribution authors
*Corresponding author*. E-mail: purbarun@iitrpr.ac.in ; pdhar1990@gmail.com
Phone: +91–1881–24–2173

## Abstract

Electrophoresis has been shown as a novel methodology to enhance heat conduction capabilities of nanocolloidal dispersions. A thoroughly designed experimental system has been envisaged to solely probe heat conduction across nanofluids by specifically eliminating the buoyancy driven convective component. Electric field is applied across the test specimen in order to induce electrophoresis in conjunction with the existing thermal gradient. It is observed that the electrophoretic drift of the nanoparticles acts as an additional thermal transport drift mechanism over and above the already existent Brownian diffusion and thermophoresis dominated thermal conduction. A scaling analysis of the thermophoretic and electrophoretic velocities from classical Huckel-Smoluchowski formalism is able to mathematically predict the thermal performance enhancement due to electrophoresis. It is also inferred that the dielectric characteristics of the particle material is the major determining component of the electrophoretic amplification of heat transfer. Influence of surfactants has also been probed into and it is observed that enhancing the stability via surface charge modulation can in fact enhance the electrophoretic drift, thereby enhancing heat transfer calibre. Also, surfactants ensure colloidal stability as well as chemical gradient induced recirculation, thus ensuring colloidal phase equilibrium and low hysteresis in spite of the directional drift in presence of electric field forcing. The findings may have potential implications in enhanced and tunable thermal management of micro-nanoscale devices.

**Keywords:** Nanofluid, thermophoresis, heat conduction, electrophoresis, smart fluid, dielectric

## 1. Introduction

Enhanced thermal conduction effect in nanocolloids/nanofluids [1–3] with respect to the base fluids has been a topic of widespread study and has been received by the academic community with equal shares of encouragement as well as scepticism. The mixed reaction has been due to the fact that the results and observations have rarely had a conclusive point of convergence and are often non–reproducible even at the laboratory scale. This is quite evident from a benchmark exercise undertaken among several research groups and there were a very few occurrences where the thermal conductivity measurements for a sample by different groups were closely similar [2]. However, there has been a consensus among a majority of academicians in the field that thermal conduction in fluids enhance by addition of nanoparticles but the degree of enhancement is often low as the concentration of particles requires to be within dilute regimes (within 1–1.5 wt. %) to ensure stability of the colloidal phase [4, 5]. As far as heat conduction is concerned, the enhancement in the transport phenomena has been explained based on several propositions. The more accepted theories for the same are enhanced transport due to Brownian flux and thermophoresis [6, 7], effective medium theory [8], aggregation pathways for conductive transport [9], nanoscale liquid layering [10] and so forth. The research community is also unanimous that only dilute nanocolloidal systems have the potential to exhibit fairly long term stability and often charged (cationic or anionic as compatible with the particle material) surfactants are employed [11, 12] to ensure extended stability of the colloidal phase.

Along with thermal properties of such colloidal media, the electrical transport characteristics is also another important aspect for nanocolloids as it is important in several applications, like enhancing breakdown voltage of liquid insulators [13, 14], modulating charge transport in bio-nano systems [15, 16], etc. The formation of Electric Double Layer (EDL) around the nanostructures in presence of a polar fluid and the formation of pseudo charge within the colloidal domain have been reported and both experimental [17, 18] as well as theoretical studies [19] have been put forward. The impact of surfactants on the EDL and the consequent enhancement in the electrical conductivity of nanocolloids has also been reported. Essentially, such colloidal systems can be considered to behave as electrically conducting media with discrete charge carriers in the form of nanostructures [19]. Hence, the electrophoretic drift component of such nanostructures can be augmented directionally by employing an external electric field. Consequently, as the nanostructures are also dynamic carriers of heat across the fluidic domain [19], directional migration of the nanoparticle population from the hot zone towards cold zones can in fact improve the thermal performance of device by augmenting thermal transport away from hot zones. The present article conceptualizes experimentally the efficacy of the idea. Induced electrophoresis is employed to improve the thermal performance of nanofluids by augmented particle migration from hot towards cold regions. The present findings clearly show that such colloids can be tuned via coupled electro-thermal effects to obtain enhanced and smart cooling of miniaturized devices.

## 2. Materials and methodologies

The experimental setup for the present studies has been designed in a manner such that the effect of electrophoresis is clearly segregated from the thermophoretic and Brownian diffusion effects. The setup consists of a fabricated test section and accessories for generation of thermal gradients and data acquisition. The schematic of the complete setup has been illustrated in Fig. 1 (a). The setup consists of a test cell which has been illustrated in Fig. 1 (b). The test cell consists of an internally hollow cylindrical nanofluid confinement, 16 mm in height, 25 mm internal diameter and 40 mm external diameter and entirely fabricated from Teflon. Teflon is chosen as the material due to its thermal and electrical insulation characteristics and its hydrophobic nature as well which prevents the nanofluid from staining the walls due to particle deposition. Both the ends of the Teflon container are open. Of the 16 mm height, the top and bottom regions contain threads machined on the inside of the Teflon container and the thread lengths are 3 mm each at the top and bottom. Two solid copper cylinders, each 8 mm in height and 12 mm in diameter, are machined and threads are machined to a length of 3 mm to match the threads of the Teflon. The two copper cylinders are threaded to the Teflon to provide the complete sealed assembly. Thin silicone gaskets are positioned between the Teflon and the copper plugs to prevent leakage. The net height of the cavity housing the nanofluid thus essentially becomes 10 mm. Teflon housing caps are fabricated and capped on to the copper plugs to insulate the whole setup.

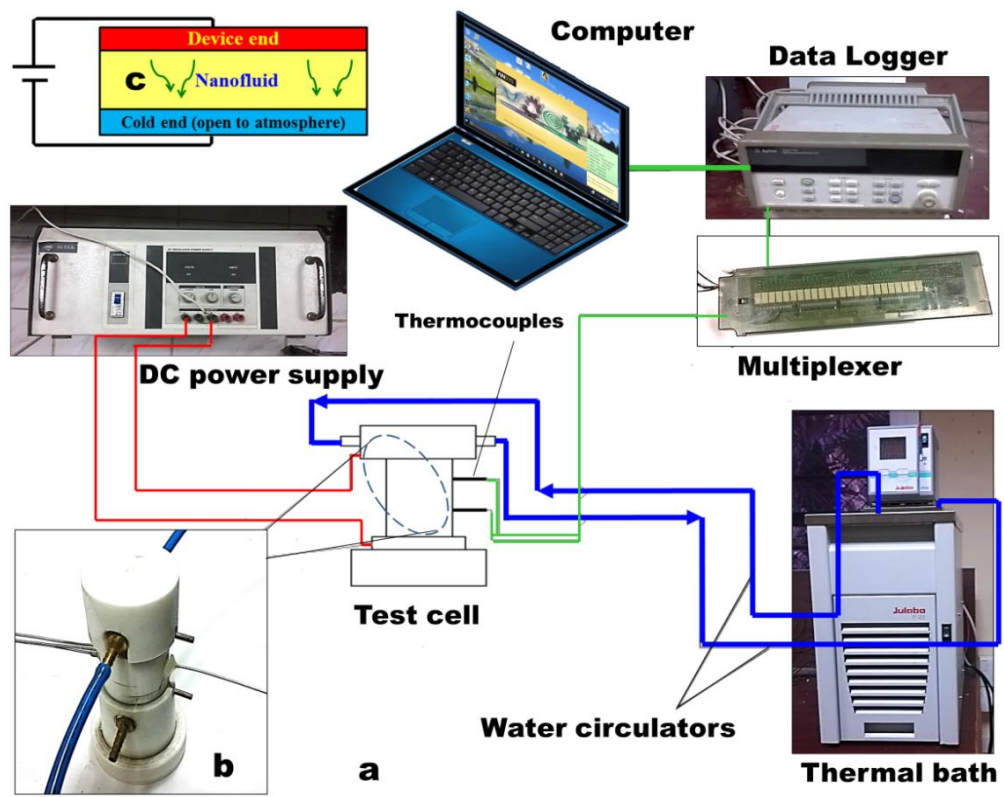

**Figure 1: (a)** Schematic of the experimental setup and the major constituents. **(b)** View of the test section with the upper piping attached for hot fluid supply and thermocouple assembly. **(c)** Illustration of the philosophy behind the present study. A small compartment of nanofluid is sandwiched between the heated device end (top) and the coolant (or ambient) end. Upon exceeding a certain temperature, the external circuit switches on, initiating the electrophoretic drift, leading to enhanced transport of heat from the hot end towards the cold end.

Each end of the central Teflon container contains 5 mm of the copper plugs buttressing out after assembly. A through hole of 2.5 mm diameter is machined through each of the buttressing end and short copper pipes are brazed to provide for fixtures to flexible water tubes (as illustrated in Fig. 1 (b)). A copper pin is also brazed to each buttressing end for clamping the crocodile clips for establishing the electrical connection. The bottom plug is screwed to the Teflon cavity and the same is filled with the required fluid and the top plug is screwed in to get the complete assembly. A minute vent hole is machined in the top plug which allows drainage of the excess fluid as the top plug thread is tightened in. After drainage of excess fluid, the vent hole is closed using a bolt using an Allen key. As observable in Fig. 1 (a), flexible tubing is used to pump water at a particular temperature through the hole in the copper plugs from a constant temperature bath. At steady state, the copper plugs attain constant temperatures and act as constant temperature boundaries across the nanofluid. Crocodile clips are attached to the pins brazed to the copper plugs and connected to the terminals of a DC power supply (Polytronic Corpn. India, 0–200 V, 0–10 A). K type thermocouples (bead diameter 0.25 mm) are attached to the system to read the temperatures using a multiplexer based data acquisition system (Keysight Inc. USA) via a computer terminal. Five thermocouples were employed and two of them were placed very close to the bottom of the top copper plug and very close to the top of the bottom copper plug respectively. The thermocouples are calibrated employing a constant temperature bath and a standard Pt-100 thermocouple and the maximum error is observed to be within ± 0.6 ºC. The contact is avoided as the copper plugs conduct electricity and will hamper the thermocouple operation. The remaining three have been placed within the fluid container with the beads positioned at the central axis of the hollow cylinder. The thermocouples have been placed at 0.25 cm, 0.5 cm and 0.75 cm from the top plate. The spacing has been kept uniform to ensure proper reading of the gradient generated within the fluid.

The philosophy behind the present work has been illustrated in Fig. 1 (c) employing a schematic. The device component which requires thermal management in the form of heat removal has been marked as device end. The heat rejection region has been marked as cold end, which may be open to atmosphere or connected to specific thermal management devices. A thin compartment flooded with the dielectric nanofluid has been sandwiched between the device and the cold ends. It is ensured that the device end, nanofluid cavity and cold end are arranged in a vertical order. An externally driven DC circuit connected to a battery system is attached across the terminals of the nanofluid chamber. During normal operation, heat is conducted by the nanofluid from the hot end and dumped at the cold end. The vertical arrangement ensures heat is conducted across the fluid. In case of enhanced heat generation by the device, the external circuit introduces an electric field across the nanofluid cell terminals. This leads to initiation of a direction electrophoretic drift from the hot end towards the cold end, essentially leading to enhanced transport of heat by the constituent nanoparticles away from the hot end. Only the top plate is maintained at higher temperature as the vertical positioning prevents any convective heat transfer and hence heat is transferred only by conduction across the fluid. This essentially provides information on the energy transport by nanofluids solely due to diffusive transport, either thermodiffusion or electrophoresis. The present concept leads to enhanced extraction of heat using smart thermal as well as tunable additional electrical effects in nanofluids.

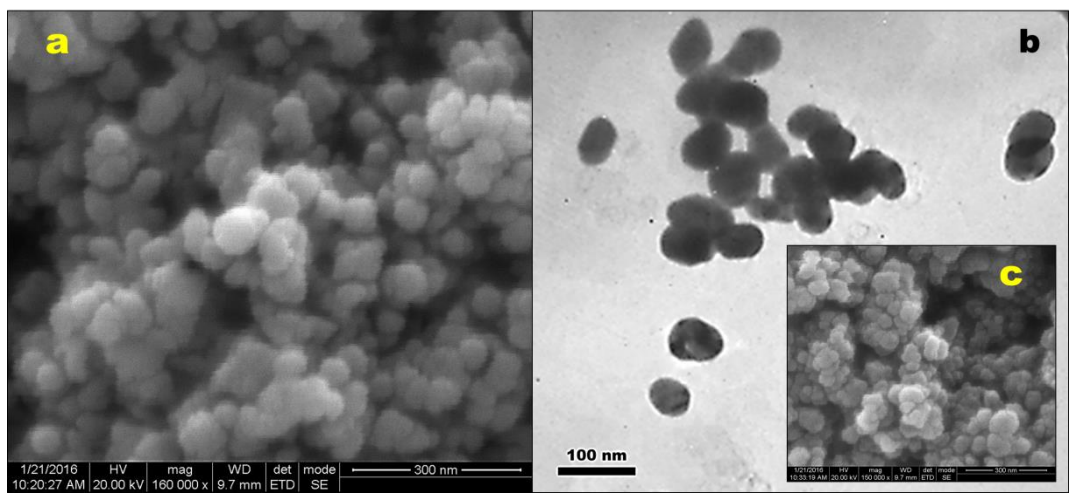

**Figure 2:** Morphology characterization of the nanostructures used in the present study **(a)** Scanning electron microscope image of copper oxide (~30–40 nm, spherical) **(b)**

Transmission electron microscope image of aluminium oxide (~50 nm, oblate) and **(c)** SEM image of titanium dioxide (~30–45 nm, distorted square cross section).

Three different types of nanoparticles have been employed for synthesis of nanofluids in the present experiments and the electron microscopy images for the same have been illustrated in Fig. 2. The nanoparticles used are copper oxide (Alfa Aeser Inc. India, ~30–40 nm spherical particles), aluminium oxide (Sigma Aldrich India, ~50 nm, oblate particles) and titanium dioxide (Nanoshel Inc., USA, ~30–45 nm particles with square/hexagonal cross section). Two different surfactants have also been employed to enhance the stability of the colloidal phase as well as increase the surface charge of the nanoparticles which in turn enhances the electrophoretic drift under the influence of electric field. Sodium dodecyl (or lauryl) sulphate (SDS) (Sigma Aldrich India) and Cetyl triammonium bromide (CTAB) (Sisco Research Labs, India) are the surfactants used in the study. Deionized water (Millipore, bulk electrical conductivity 2–3 μS/cm, measured using Cyberscan probe, USA, thermal conductivity 0.58 W/mK, measured using KD2 Pro, Decagon devices, USA) has been used as the base fluid. Nanocolloids were synthesized by dispersing the required amount of nanoparticle and surfactant in the base fluid and stabilized employing a probe ultrasonicator (Oscar Ultrasonics, India). The resultant fluids were observed to exhibit shelf life stability of 5–20 days depending on the colloid prepared. The electric field strength has been restricted to 60 V/cm as beyond that very low enhancement is observed in heat transfer as the enhanced electrophoresis leads to permanent migration of the nanoparticles, leading to decreased colloidal stability and enhanced chances of phase separation near the copper plug electrode terminals.

## 3. Results and discussions

In order to determine the effects of electrophoretic drift on the thermal conduction calibre of nanofluids, the experiments were performed in a simple manner. Initially, the effect of particles on the thermal transport effect was determined. The top copper hub was fed with water at high temperature (at 60 ± 0.5 °C) while the bottom plate was kept open to the ambient by clamping the whole test section across its middle and allowing it to remain freely suspended in air. The ambient temperature was maintained constant employing an air conditioned environment (at 20 ± 0.5 °C, as obtained from temperature measurement of the ambient just near the bottom copper plug) employing air conditioning of the experimental region. In essence, the fluid within the confinement (water) is extracting heat from the heated top plate and releasing the same at the cold bottom plate, solely by conduction as convection has been arrested by design by preventing a buoyant gradient to exist. The evolution of temperature at the bottom plate is registered as time progresses and the net thermal gradient across the test cell is calculated with progressing time. Since the total height of the confined fluid body is 1 cm, the magnitude of the net gradient becomes equal to the difference in temperature between the top and bottom plates when expressed in °C/cm. To initiate the exercise, steady state thermal gradient generation for water and stable nanofluids (without field) are observed as control experiments and have been illustrated in Fig. 3 (a).

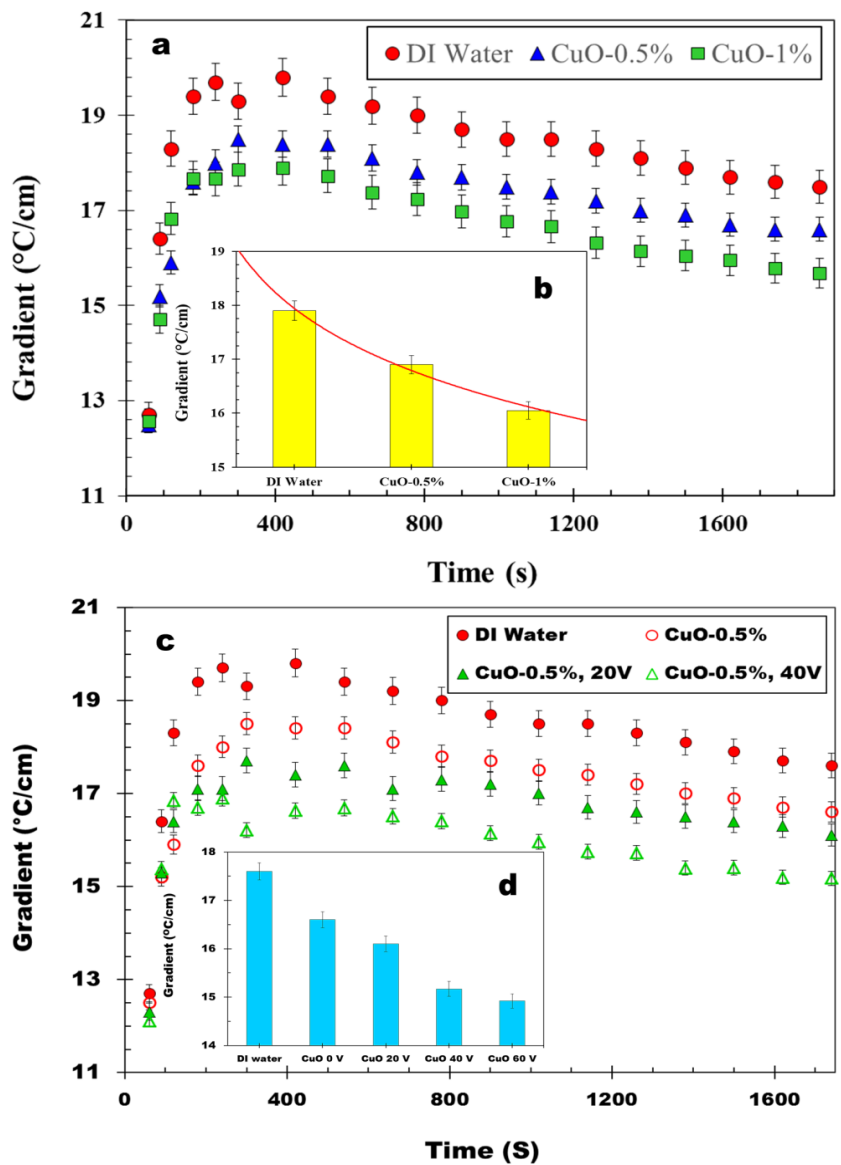

**Figure 3:** (a) Gradient generation and temporal evolution comparison in case of water, 0.5 wt. % and 1 wt. % CuO–water nanofluids for same hot end temperature and ambient conditions. (b) Steady state temperature gradient magnitude across the fluid for different fluids. (c) Gradient generation and temporal evolution comparison in case of water and CuO–water nanofluids with applied electric field across the fluid and conditions same as applied in case (a). (d) Steady state temperature gradient magnitude across the fluid for different field intensities.

As observable from Fig. 3 (a), the magnitude of the steady state gradient in case of the nanofluids is appreciably reduced when compared with respect to water and the reduction enhances with increasing nanoparticle concentration. The gradient is calculated based on the differences in temperatures at the top and bottom plates. Assuming Fourier conduction across the fluid during steady state from the hot top plate to the bottom cold plate, the heat flux $q''$ can be expressed as a function of thermal gradient along cylinder depth $z$ and the effective thermal conductivity of the fluid $k_f$ as

$$q'' = -k_f \frac{T_H - T_C}{z} \qquad (1)$$

As the temperature of the top plate is constant, the temperature of the bottom plate at steady state is dependent on the thermal conductivity of the fluid and the heat flux transferred across the fluid. Since the heat sources are equal in all cases (obtained by flow of circulating water through the top copper hub), the heat flux to which the inner fluid is subjected to is equal. Since the measured value of the thermal gradient is lower for the nanofluid compared to water, it thereby signifies that the thermal conductivity of the conducting fluid media has enhanced. This is a very well established and well documented phenomena that thermal conductivity of nanofluids is augmented compared to the base fluid and essentially the observation in Fig. 3 (a) also act as a validation study for the experimental facility and its accuracy. The magnitude of the steady state temperature gradient has been illustrated for the fluids in fig. 3 (b). it may be noted that a reduction of ~ 1.5 °C/ cm is obtained for a 1 wt. % CuO nanofluid compared to the case of water, which in the present case signifies increase in the bottom plate steady state temperature by ~ 1.5 °C. This observation validates that thermophoresis and collisional transport of heat by particle collisions across the nanofluid is able to extract more heat from the top plate compared to the base fluid and dump the same at the bottom plate.

Having established the efficacy of nanofluids in enhancing heat conduction across the fluidic chamber from the top hot zone to the bottom cold zone open to ambient, the influence of applied electric field is next probed. The same configuration of constant temperature top plate and bottom plate open to ambient has been maintained with the addition of small magnitudes of DC voltage applied across the terminals of the top and bottom plates. Initially, 20 V is applied to one of the plates and the other plate is maintained at ground potential, creating a field measuring 20 V/cm across the nanofluid. The fluid is then allowed to reach a thermally steady state for the same top plate fixed temperature. The observations have been illustrated in Fig. 3 (c) for a water based CuO nanofluid of 0.5 wt. % concentration. It is observed that increasing the applied electric field leads to further decrease in the steady state temperature gradient value which essentially signifies that the bottom plate has gained higher temperature than the zero field case. Such an observation reveals that an additional mechanism is at play which ensures higher heat extraction from the top plate and enhanced release at the cold bottom plate. It may be noted that since the nanofluids have weak bulk electrical conductivity (the 0.5 wt. % CuO–water nanofluid has a measured bulk electrical conductivity of ~100 μS/cm for average CuO particle size of 30 nm), the current traversing across the fluid from the top to the bottom plate is negligibly small and hence Joule heating of the plates can be ruled out as a possibility for observed reduced thermal gradient.

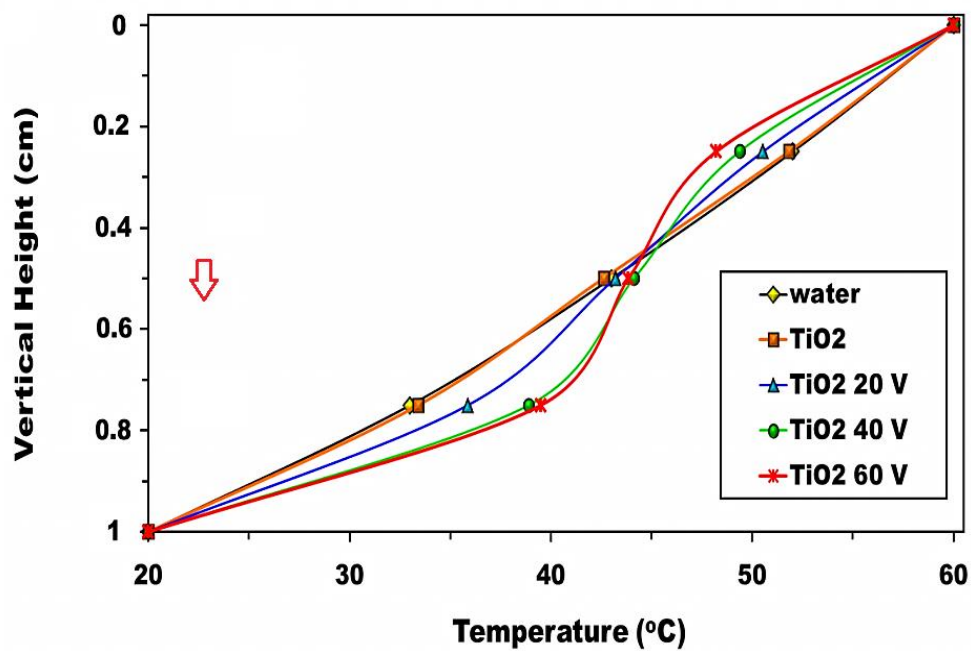

**Figure 4:** Transformation of the shape of temperature profile within the nanofluid from linear to a sigmoid pattern exhibiting extraction of heat from hot zone and deposition to the cold zone by electrophoretic migration. A minute variation from the linear profile in case of nanofluid without field could be attributed to thermophoretic drift. The arrow shows the direction of the bottom plate, with vertical zero representing the top plate.

It is observable from fig. 3 (d) that employing an electric field of 60 V/cm (a very moderate field), the thermal gradient can be further lowered by ~1.7 °C/cm compared to the simple nanofluid, which signifies further increment in the heat conduction effect. Whenever an interface is created in conjunction with a polar fluid, an electric double layer (EDL) is formed at the interface (in this case the nanoparticle–fluid interface) due to net dominance of a particular charge at the interface. The dominant charge forms a layer (Stern-Helmholtz layer) closely packed to the interface, giving rise to a surface potential termed as the Stern potential. The non-dominant charge distributes around the Stern layer as the Guoy-Chapman layer and the two layers constitute the EDL. The Chapman layer contains both the charges distributed as per the Poisson-Boltzmann distribution function and diffuses off at a particular distance, beyond which the effect of the EDL decays out. This distance is called the Debye thickness ($\lambda$) and is expressible as

$$\lambda = \sqrt{\frac{\varepsilon_0 \varepsilon_r k_B T}{2 n e^2 z^2}} \qquad (2)$$

Where '$\varepsilon$' represents the dielectric permittivity (subscripts '$0$' and '$r$' represent free space and relative permittivity for the nanofluid), $k_B$, $T$, $n$, $e$ and $z$ represent the Boltzmann constant, absolute temperature, number density of ionic population, elementary charge and valence of the ionic components respectively. In a nanoparticle based colloidal system, the EDL is expected to be fairly thin compared to the particle diameter as particles are uncharged entities and the EDL only forms due to interactions of the surface atoms with the neighbouring polar water molecules. Consequently, the associated Dukhin number for the system is small. The external electric field thus causes the interfacially charged particles to possess a directional electrophoretic velocity ($v_{el}$), the magnitude of which can be expressed by the Huckel-Smoluchowski equation for small Dukhin number systems as

$$v_{el} = \frac{\varepsilon_0 \varepsilon_r \zeta}{\mu} E \qquad (3)$$

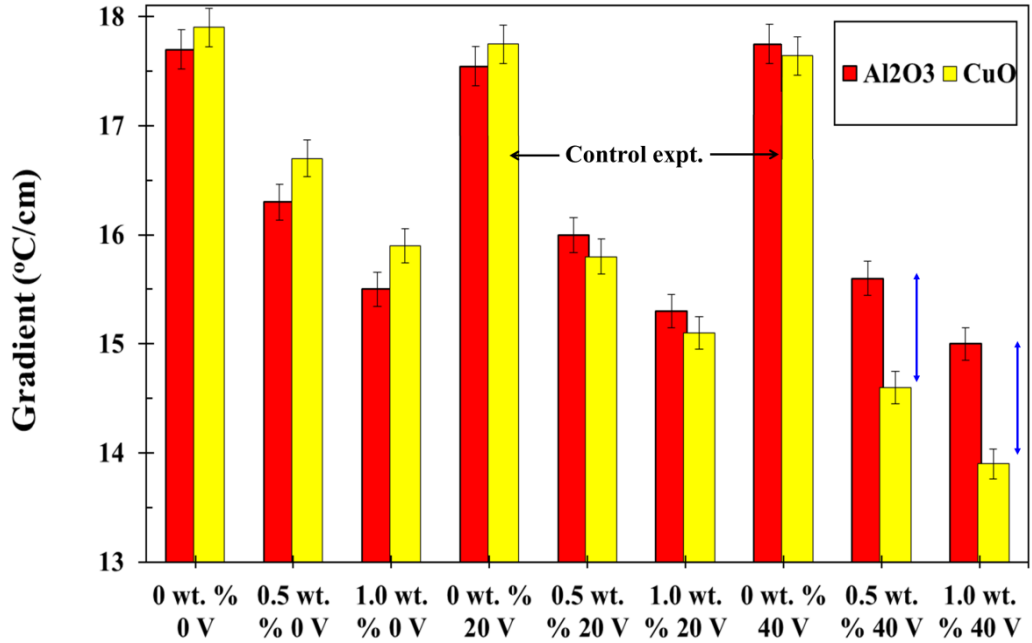

**Figure 5:** Aggregate of effects of nanoparticle dielectric constant, thermal conductivity, concentration and field strength on the enhanced thermal conduction within the nanofluid as manifested through the reduced thermal gradient across the fluid.

In the absence of field, the thermal gradient at steady state is reduced compared to the base fluid due to thermal migration of the particles from hotter regions to colder regions, thereby transferring momentum and energy towards the cold plate in a cascaded mechanism. This migration is termed as thermophoresis and maybe deemed responsible for increased thermal transport in case of the nanofluid. Since the present experiment induces a directional thermal gradient, the impact of Brownian diffusion in the heat transfer process maybe neglected due to its random fluctuating spectra. For nanoparticles stabilized in a fluid phase, the thermophoretic velocity ($v_{th}$) is expressible as

$$v_{th} = \frac{-0.55 \mu \nabla T}{\rho T} \qquad (4)$$

Considering a generic nanoparticle of CuO of average particle diameter of 30 nm dispersed in water, a scaled thermophoretic velocity may be obtained as ~1.752 nm/s. For the scaling analysis, the average temperature is assumed to be average of the top and bottom plates and the calculations have been performed for unit local thermal gradient as determining the exact scale for the localized thermal gradient at the nanoparticle-fluid interface is a challenging task. Hence the additional directional flux per unit thermal gradient leads to enhanced conduction via collisional energy transport across from hotter zones to colder zones, leading to reduction in the steady state thermal gradient. A generic CuO-water colloid with 0.5 wt. % nanoparticle concentration has been tested to obtain a zeta potential value of ~45 mV. Accordingly at 60V/cm field intensity, the electrophoretic velocity is obtained to be ~3.5 nm/s. Consequently, the net directional drift velocity of the particles enhances, causing further collisional transport of heat. Hence, a reduced order simplistic scaling of the drift velocities is able to explain the enhanced thermal transport by induced electrophoresis.

Further experimental evidence of enhanced thermal conduction away from the hot plate towards the colder plate can be provided by fixing the thermal boundary conditions and probing the nature of the thermal gradient within the fluid. The top plate is maintained at 60 °C and the bottom plate is now maintained at 20 °C by passage of conditioning water from two independent temperature control baths. At steady state, a thermal gradient which is linear in nature is obtained as illustrated in Fig. 4. The temperature values are obtained at 5 locations, viz. the top plate, 0.25 mm below the top plate, 0.5 mm below, 0.75 mm below and 1 mm below, i.e. at the bottom plate employing the thermocouples. In case of a 1 wt. % $TiO_2$ based aqueous nanofluid, the linear profile is observed to be maintained at steady state as any form of non-equilibrium thermophoretic drift has died out once a steady thermal gradient is reached. Now, the electric field is applied for 300 seconds and the temperature profiles are recorded. It is observed that the electrophoretic drift leads to extraction of heat from the hotter end of the contraption and the heat is transported towards the colder end. In the present case as the bottom plate temperature is fixed, the removal of heat to atmosphere is less compared to the case where the bottom place is open to atmospheric convection. This causes increment in temperatures near the bottom plate and reduction near the top plate, giving rise to a non-linear, sigmoid thermal distribution deviating from the linear profile.

It may be further mentioned here that Joule heating within the nanofluid is not responsible for the disruption in the steady state profile as it is not possible to obtain simultaneous decrease and increase in temperatures in case of Joule heating.

The discussion so far reveals that the electrophoretic mobility is an essential determining factor for the performance of the nanofluid as a smart fluid. This essentially indicates that the size of the particles as well as the dielectric characteristics of the particle material is important rate determining features. In case of similar sized particles, the relative permittivity determines the ability of the nanofluid as electrophoretic material. Fig. 5 represents a similar control experiment wherein similar sized particles are employed but different particle materials are chosen to portray the importance of the dielectric constant. It involves aluminium oxide (~50 nm) and copper oxide (~ 30 nm-40 nm) based nanofluids under the influence of electric field. The control experiments in absence of field illustrates that both the systems exhibit similar values of steady state thermal gradients and thermophoretic transport in a strong function of particle size and localized non-equilibrium thermal gradient and thereby the different thermal conductivities of the two materials have minimal impact. However, in presence of electric field, the difference is evident. At 40 V, the improved performance of the copper oxide (relative permittivity ~ 18) is explicitly observed over the aluminium oxide relative permittivity ~ 10), which from the electrophoretic drift equation essentially signifies that the dielectric permittivity of the materials. It is of course needless to mention that the zeta potential will be a function of the dielectric permittivity, however, given that both dielectric permittivity as well as zeta potential are present in the numerator of the Huckel equation (Eqn. 3), the net influence of dielectric permittivity is twin fold. This is further evident from the fact that the electrophoretic drift induced heat transfer is largely amplified for copper oxide at 40 V compared to 20 V in Fig. 5.

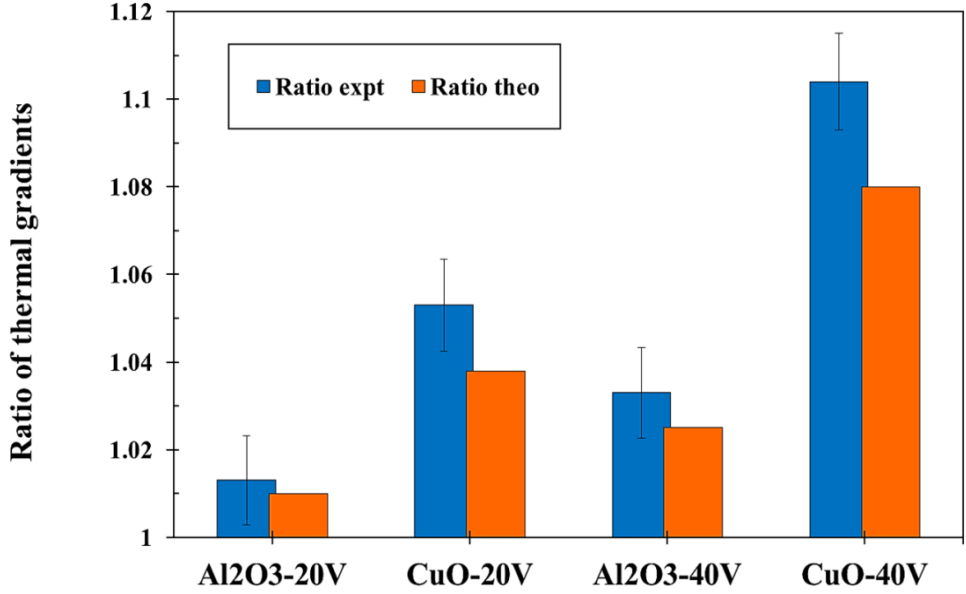

**Figure 6:** Comparison of the experimental observations with the scaling analysis for predicting the enhancement in thermal performance due to electrophoretic drift.

Keeping in mind the scaling analysis discussed earlier to explain the phenomena of enhanced heat conduction due to electrophoretic drift over thermophoresis, the same analysis maybe extended to model the phenomena with appreciable accuracy. The energy transported by virtue of velocity of suspended particles maybe scaled in equivalence to the kinetic theory of gases, wherein the average energy transported scales directly to the average drift velocity of the particle. Thereby, for the simple case in absence of field, the energy transported by a particle would scale to the thermophoretic drift velocity. Similarly, in case of added field, the transported energy would scale as the algebraic sum of the thermophoretic velocity and the electrophoretic velocity. The algebraic sum would suffice based on the argument that both the thermophoresis and electrophoresis are co-directional due to the geometric implications of the designed experimental test assembly. The effective energy transported would lead to a final thermal gradient value, which from Fourier's law of conduction would be inversely scaled to the net energy transported due to particle drift (higher energy transport leads to lower value of thermal gradient across the system). Based on the above discussion, the net thermal gradient maybe described mathematically as

$$\frac{\left(\frac{dT}{dz}\right)_{f0}}{\left(\frac{dT}{dz}\right)_{f}} \propto \left(1 + \frac{v_{el}}{v_{th}}\right) \quad (5)$$

Here subscripts 'f' and 'f0' indicate with field and without field. Accordingly, the ratio of thermal gradients for field on condition and no field conditions may be evaluated experimentally for different nanofluids and the same may also be theoretically deduced via reduced order scaling analysis. Accordingly, representative values of the deduced thermal gradient ratio for various conditions have been illustrated against experimentally determined values in Fig. 6 and it is observed that the first principles scaling model may be employed to predict the electrophoretic heat conduction augmentation in nanofluids with appreciable accuracy.

Influence of external electric fields may in certain cases lead to phase separation of the colloid due to excessive dielectrophoretic drift of the particles with respect to the fluid field such that conditions prerequisite localized sedimentation maybe achieved. In order to improve stability of the particle phase in suspension, suitable surfactants are employed. The usage of ionic surfactants served a twofold purpose, viz. increment in the stability of the colloid to ensure no phase separation during electrophoresis and the enhanced interfacial charge due to the charged surfactant molecules leads to augmented zeta potential which increments the electrophoretic mobility, thereby directly enhancing the thermal conduction process further. Furthermore, usage of surfactants also allows employing nanomaterials of high dielectric constant such as titanium dioxide, which otherwise forms a very unstable colloidal phase in polar solvents such as water. The use of stabilized nanomaterials of high dielectric constant enables further augmented electrophoresis induced heat conduction without hampering colloidal phase stability. The influence of $TiO_2$ (30-45 nm, rutile phase) based aqueous nanofluids (stabilized with 4 mM Sodium Dodecyl Sulphate (SDS)) on electrophoresis induced thermal transport has been illustrated in Fig. 7 (a). The surfactant concentration has been chosen so as to keep it at half of its Critical Micelle Concentration (CMC) so as to effectively enhance colloidal phase stability yet prevent micelle formation by the surfactant chains, leading to excessive frothing and hampering of thermal transport across the suspensions. It may be evidently observed that while copper oxide was able to lower the thermal gradient across the test cell to values of 16 $^{\circ}$C at best from the original value of 19 $^{\circ}$C for an electric field intensity of 40 V/cm, the stabilized titania based nanofluid is able to lower it much further and near to 12 $^{\circ}$C. The same is possible due to the high dielectric constant of titania (~ 120) compared to copper oxide. It may be noted that despite the similar density and lower thermal conductivity of titania compared to copper oxide nd aluminium oxide, it is still able to influence the amplified conduction process with greater efficacy, thereby further cementing the deduction that the dielectric character of the particle is at the crux of the mechanism.

However, it is also noteworthy that the presence of the surfactant also aids the electrophoresis since they themselves are charged entities and act actively towards modifying the EDL of the particle-fluid interface. Accordingly, a segregation exercise yielding clarity on the specific contribution by particles and surfactants on the electrophoretic heat transfer augmentation physics is necessary. Fig. 7 (b) illustrates the influence of surfactant (CTAB) concentration for a constant particle concentration (0.5 wt. %) on the electrophoretic thermal transport. It is observed that with increment in the value of induced voltage, the thermal performance of 1 CMC surfactant amplifies with respect to 0.5 CMC surfactant concentrations. Fig. 7 (c) illustrates the influence of particle concentration for a constant surfactant concentration (1 SDS). It is observed that increment of particle concentration leads to amplified thermal transport compared to the dilute colloidal system. However, the improvement in heat conduction calibre is observed to be higher in case of constant surfactant case than the constant particle concentration case. The analysis thus cements the ideology that electrophoresis by the particles has greater influence on the heat conduction and that the presence of surfactants simply enhances the influence of the EDL and augments the induced electrophoresis.

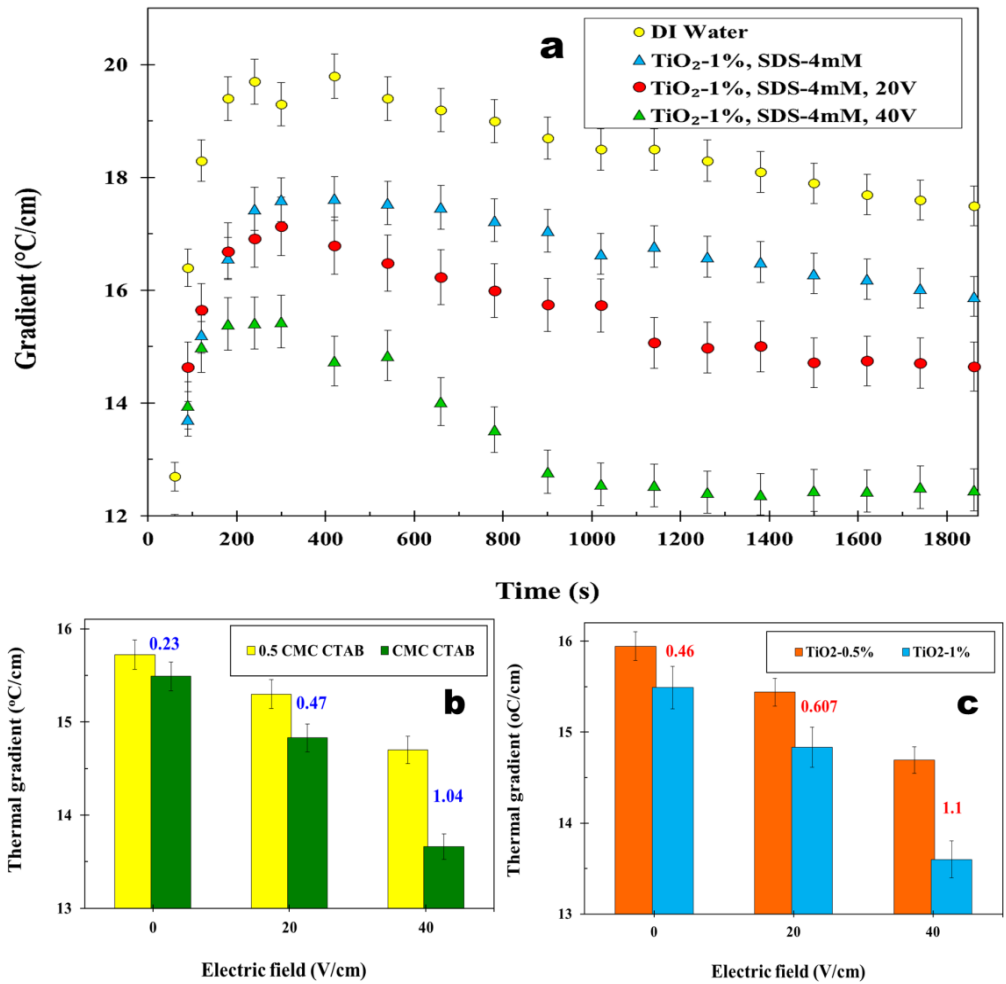

**Figure 7:** (a) Enhanced electrophoresis induced thermal transport in case of high dielectric constant particles (TiO$_2$) as well as with addition of ionic surfactants (b) Segregated contribution towards electrophoresis induced heat transfer due to surfactant concentration and (c) due to particle concentration. The numbers in (b) and (c) indicate the differences in magnitudes of two annexe columns.

Having discussed the potential of electrophoresis induced enhanced heat conduction potential of nanofluids; it is also pertinent from an engineering point of view to probe into the hysteresis in heat conduction in case of a transient electric field. The hysteresis study is important from two fronts, viz. it allows to understand the response of a particular nanofluid to a particular field strength and it provides a clear picture of the loss in heat conduction calibre (if any) on repeated on–off cycles of the electric field. Fig. 8 (a) illustrates the effective hysteresis behaviour for a 1 wt. % CuO nanofluid, as noted at two different thermocouples. It may be noted that the effective hysteresis is relatively higher in case of higher applied field strength. A stronger field leads to higher degree of electrophoretic drift and hence the localized non-equilibrium induced within the concentration profile of the nanofluid is much more prominent. Accordingly, the species diffusion cycle which restores the local concentration equilibrium upon field withdrawal requires additional effort and time to negate the created concentration gradient. Additional time is needed as the diffusion constant for the nanofluid remains same irrespective of the field strength applied and hence higher concentration non-equilibrium calls for longer relaxation time. Consequently the hysteresis in conduction in the next field cycle is higher due to remnant concentration non-equilibrium and observations in Fig. 8 (a) are thus consistent.

It is further observed that for the same electrophoretic strength, the hysteresis at the zone of higher temperature is more than that in case of lower temperature. At higher temperatures, the strength of the Brownian fluctuation is higher compared to that at lower temperatures. Since the diffusive transport due to electrophoresis is already present, the thermal gradients near the hot zone is higher than the lower temperature zone, thus leading to stronger thermophoretic flux also at the higher temperature zone. The high degree of thermal fluctuation ensures that the concentration gradient engine trying to bring about localized equilibrium post field withdrawal is hampered to a certain extent, leading to hysteresis. Fig. 8 (b) illustrates the effect of particle dielectric constant on the heat extraction calibre as well as induced hysteresis. Three ceramic particles of nearly equivalent diameter (~35 nm) has been considered, viz. aluminium oxide (k ~ 30 W/mK, $\varepsilon_r$ ~ 10), copper oxide (k ~ 20 W/mK, $\varepsilon_r$ ~ 20) and titanium dioxide (k~ 10 W/mK, $\varepsilon_r$ ~ 120). As the size distribution Gaussians nearly overlap for the three particles, the influence of size maybe considered negligible in the present case. As the dielectric constants of the particles are widely ranged, a clear effect of the same on electrophoretic thermal extraction can be presented. The effect of thermal conductivity of the particle can also be understood from the three cases. Given the fact that the fluidic compartment is closed at both ends, it poses a constant mass constraint. Accordingly, the electrophoretic drift of the particles creates fluid motion due to drag. To satisfy the closed mass constraint, regions of backflow should also exist within the system which causes mixing and back diffusion of the particles towards their initial positions. Furthermore, in surfactant based systems, a chemical gradient is also generated due to drift of surfactant laden nanoparticles away from the hot plate due to electrophoresis.

This also drives a chemical diffusion engine from the cold towards the hot plate which causes back migration of the surfactant laden particles, thereby equilibrating the nanoparticle concentration deficit near the hot plate. Accordingly at such weak electric field intensities, permanent drift of the particles that might cause colloidal phase separation is not observed and thus forms a low hysteresis and reusable system.

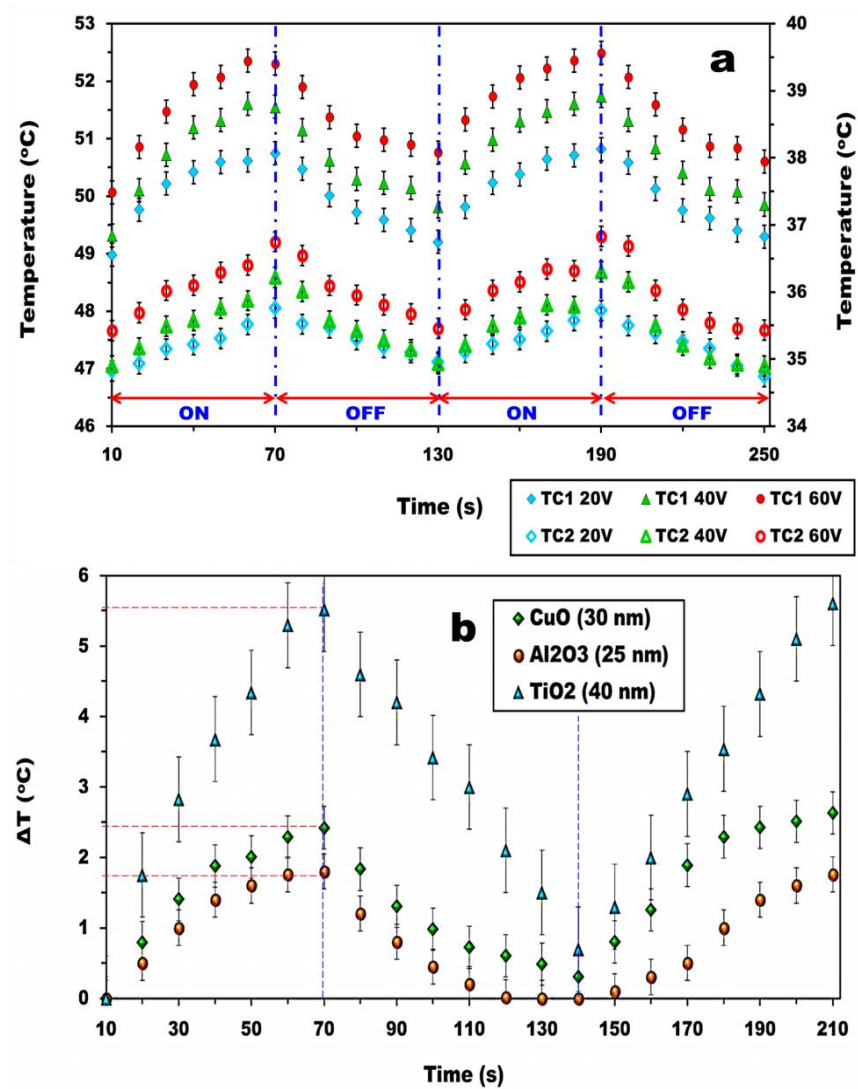

**Figure 8:** (a) Thermal hysteresis in the nanofluid sample due to cyclic on and off states of the induced electrophoresis exhibits very low lag effect. Thermocouple (TC) 1 data is plotted on left hand y-axis while TC2 is on the right hand y-axis. (b) Performance comparison for different nanomaterial based nanofluids based on the maximum temperature rise observed at the bottom plate due to enhanced thermal migration at 1 wt. % concentration and 60 V forcing field strength. The $TiO_2$ sample is observed to achieve a very high heat extraction capacity.

## 4. Conclusions

To infer, electrophoresis has been shown as a novel methodology to enhance heat conduction capabilities of nanocolloidal dispersions. A thoroughly designed experimental system has been envisaged to probe heat conduction across nanofluids by specifically removing the natural convective component. Design has been incorporated to observe the thermal history and distribution within the fluid at various time and spatial locations. Voltage is applied across the test specimen in order to induce electrophoresis in conjunction with the existing thermal gradient. It is observed that the electrophoretic drift of the nanoparticles acts as an additional thermal transport drift mechanism over and above the already existent Brownian diffusion and thermophoresis dominated thermal conduction. A scaling analysis of the thermophoretic and electrophoretic velocities from classical Huckel-Smoluchowski formalism is able to mathematically predict the thermal performance enhancement due to electrophoresis. It is also inferred that the dielectric characteristics of the particle material is the major determining component of the electrophoretic amplification of heat transfer. Influence of surfactants has also been probed into and it is observed that enhancing the stability via surface charge modulation can in fact enhance the electrophoretic drift, thereby enhancing heat transfer calibre. Also, surfactants ensure colloidal stability as well as chemical gradient induced recirculation, thus ensuring colloidal phase equilibrium and low hysteresis in spite of the directional drift in presence of electric field forcing. The present letter describes and experimentally demonstrates a novel methodology of enhancing heat conduction in nanocolloidal systems by inducing moderate electric field across the same. The findings may find potential implications in enhanced thermal management of micro-nanoscale devices.

## Acknowledgement

PD acknowledges partial financial support by Department of Mechanical Engineering, IIT Ropar.